\documentstyle[aps,prd]{revtex}
\input amssym.def
\input amssym.tex
\def\OP{\mathop{\bigoplus}\limits}
\begin{document}

\title{On the superselection sectors of fermions}

\author{Andrew Chamblin}

\address {\qquad\\ DAMTP, Silver Street\\
Cambridge, CB3 9EW, England
}

\maketitle

\begin{abstract}

{\small We classify elementary particles according to their behaviour under the
action of the full inhomogeneous Lorentz group.  For fundamental fermions, this
approach leads us to delineate fermions into eight basic families or `types',
corresponding to the eight simply connected double covering groups of the 
inhomogeneous Lorentz group (the `pin' groups).  Given this classification, 
it is natural
to ask whether or not fermion type determines a superselection rule.  It is
also important to determine what observable effects fermion type might have;
for example, can the type of a given fermion be determined by laboratory 
experiments?  We address these questions by arguing that if multiple fermion
types really did occur in nature, then it would be mathematically equivalent and also
much simpler to think of the different types as being different states
of a {\it single} particle, which would be a particle which lived in the direct
sum of Hilbert spaces associated with the different particle types.  In the language of
group theory, these are pinor supermultiplets.  We discuss the possible 
experimental ramifications of this proposal.  In particular, following work of
J. Giesen, we show that the symmetries of the
electric dipole moment of a particle would be definitely affected by this proposal.  
In fact, we show that it would be possible to use the electric dipole moment of a
particle to determine the type.  We also present an argument that M-theory
may provide the mechanism which selects a {\it unique} pin bundle.}\\

\end{abstract}

{\noindent \bf 1. Introduction}\\

{The idea of a `superselection rule' in quantum mechanics has a long and distinguished
history [1]. In general, such a rule allows one to decompose some Hilbert space of
states, ${\cal H}$, into a direct sum of subspaces ${\cal H}_{i}$ (called
`superselection sectors'): ${\cal H} = \OP_{i}{\cal H}_{i}$, such
that the superposition principle holds in each ${\cal H}_{i}$, but such that a linear
combination, ${\alpha}{\psi}_{1} \,+\, {\beta}{\psi}_{2}$, of states ${\psi}_{1}$ and
${\psi}_{2}$ from distinct superselection sectors is not physically realisable, except
as a mixture with density matrix}
\[
|{\alpha}|^{2}\,{\psi}_{1} \,\otimes\, {\overline {\psi}_{1}} \,+\, 
|{\beta}|^{2}\,{\psi}_{2} \,\otimes\, {\overline {\psi}_{2}}
\]

{A simple example of an observable which determines a superselection rule is given by
the operator $(-1)^{F}$, which is even for states of integer spin (bosons) and odd for
states of half-integer spin (fermions). Clearly, given a fermionic state ${\psi}_{f}$
and a bosonic state ${\psi}_{b}$, we can assign no physical meaning to the linear
combination ${\alpha}{\psi}_{f} \,+\, {\beta}{\psi}_{b}$. For consider the action of
$R_{2\pi}$ (rotation in space through $2\pi$ about any axis) on such a state:}
\[
R_{2\pi}({\alpha}{\psi}_{f} \,+\, {\beta}{\psi}_{b}) = -\,{\alpha}{\psi}_{f} 
\,+\, {\beta}{\psi}_{b}
\]
{Since $R_{2\pi}$ must map any physical state to an indistinguishable state, it
follows that we must take ${\alpha} = 0$ or ${\beta} = 0$, i.e., it is impossible to
superimpose bosons and fermions.

In this paper, we address the issue of whether or not it is possible to define
superselection sectors of fermions in terms of the definitions of discrete
transformations such as $P$ and $T$. More precisely, there is always some ambiguity in
how one defines $P$ and $T$ corresponding to the ambiguity in sign: $P^{2} = \pm 1$,
$T^{2} = \pm 1$, $(PT)^{2} = \pm 1$. Traditionally, it has been argued that a choice of
signs for $P^{2}$, $T^{2}$, and $(PT)^{2}$ determines a distinct superselection
sector of fermions, each sector corresponding to a different `type' of elementary
particle. Here, we discuss how one might go about forming coherent
superpositions of fermionic states of different `type'. This is achieved through a new
construction of `type-doubling', i.e., increasing the dimensions of the fermions to
accomodate different type states simultaneously. Each fermion type is then just
a state in a higher dimensional multiplet.  Before proceeding with this
construction, however, it is useful to review some basic mathematical facts and
terminology.}\\
\vspace*{0.6cm}

{\noindent \bf 2. Fermions, Pin Groups, and Discrete Transformations}\\

{The study of fermions begins with the Dirac equation:}
\begin{equation}
{(i{\gamma}^{\mu}\,{\partial}_{\mu} - m)\,{\psi} = 0}
\end{equation}
{Dirac derived (1) by taking the square root of the standard
relativistic energy-momentum relation, and making the canonical substitutions
of momenta
for differential operators: $p_{\mu} \,\rightarrow \, i\,{\partial}_{\mu}$.
Dirac found
that the equation could only be satisfied if the ${\gamma}^{\mu}$s were
actually $4
\times 4$ {\it matrices} satisfying precisely the Clifford algebra relation:}
\[
\{{\gamma}^{\mu}, {\gamma}^{\nu}\} = 2g^{\mu \nu}
\]
{where $g^{\mu \nu}$ was (for Dirac) the flat Minkowski space metric. Thus, the
actual
wavefunction ${\psi}$ representing the electron is a {\it four-component}
object and
we are led naturally to the concept of antiparticles.

Once we form the set of solutions to equation (1) (and put an inner product
structure `$<,>$'
on that space so that it becomes a Hilbert space, denoted ${\cal H}$), it is
natural
to consider the representation of discrete geometrical transformations on
${\cal H}$.
Because the nature of these representations, in their most general form, is a
core
issue in this paper, we feel it is probably useful to include a brief
digression on
the representations of a group on a vector space (in this case, a Hilbert
space).
To this end, let $(M, g)$ be our underlying spacetime manifold, and let $A$ be
some
group of coordinate transformations on $(M, g)$. This group
could be some global group of isometries (if the manifold admits
a circle action for example) or it could be the local orthogonal group
induced pointwise by the metric structure.  Suppose that there exists a
collection of maps, ${\{}O(a_{i})|{\forall}~a_{i} ~{\in}~ A{\}}$, with the
property
that at each $a_{i} ~{\in}~ A$, {\it $O(a_{i})$ is a linear operator on some
Hilbert
space ${\cal H}$.} Then we say that the collection of linear maps $O(A) =
{\{}O(a_{i})|a_{i} ~{\in}~ A{\}}$ forms a {\it representation} of the group $A$
on the
Hilbert space ${\cal H}$ if the group structure is preserved, i.e., if
$O(a_{i})\,O(a_{j}) = O(a_{i}\,a_{j})$, for all $a_{i}$, $a_{j} ~{\in}~ A$. Such
a
representation is said to be {\it unitary} if the corresponding maps $O(a_{i})$
are
unitary operators on ${\cal H}$. A subspace ${\cal H}_{1} ~{\subset}~ {\cal H}$
is
called {\it invariant} if ${\forall}~v ~{\in}~ {\cal H}_{1}$, $O(a_{i})\,v
{}~{\in}~
{\cal H}_{1}$ for any $a_{i} ~{\in}~ A$. A representation is also said to be
{\it
reducible} if there exists an invariant subspace ${\cal H}_{1} {\not=} {\cal
H}$ whose
orthogonal complement ${\cal H}^{\perp}$ is also invariant. Otherwise, the
representation is said to be {\it irreducible}.

Of course, a set of linear operators on a vector space is itself often a
vector
space. We can therefore talk about 'representing' the geometrical symmetries of
$A$ on
the space $M({\cal H})$ = ``the set of all linear operators on ${\cal H}$".
Clearly,
$M({\cal H})$ contains all of the observables in our theory. For example, let
$H$ denote a time-independent Hamiltonian. Then we say that a geometrical
transformation $a ~{\in}~ A$ is a {\it symmetry} if $O(a)\,H\,(O(a))^{-1} = H$,
i.e.
if the two linear operators $O(a)$ and $H$ commute.

Now, in this paper we are going to introduce operators which are not
unitary; in fact, we are going to follow Wigner [17] and represent time 
reversal as an {\it anti-unitary} operator.  Recall that an operator $O$
is defined to be anti-unitary and antilinear if for any two states $\phi$ and $\psi$ of 
the system
\[
<O{\phi} \mid O{\psi}> ~=~ <{\phi} \mid {\psi}>^{*} ~=~ <{\psi} \mid {\phi}>
\]
and
\[
O\left(a\mid {\phi}> \,+\: b\mid {\psi}>\right) ~=~ a^{*}O\mid {\phi}> \,+\: 
b^{*}O\mid {\psi}>
\]

Ordinarily, the time reversal operator is chosen to be anti-unitary in order to insure
that positive energy states are mapped to positive energy states.  Since the 
product of a unitary operator and an anti-unitary operator is anti-unitary,
and parity inversion is unitary, it follows that the combined operation of 
parity inversion with time reversal is anti-unitary.  This state of affairs will
hold for all of the operators which we write down in this paper, i.e., time
reversal and the combined operation of parity inversion with time reversal will
always be anti-unitary.  This choice is the standard choice made in the particle
physics literature; DeWitt-Morette et al [8] refer to this choice as the
physical or `non-relativistic' choice.  In many books, a representation is
{\it defined} to be a representation of a group by {\it unitary} operators.
Thus, in this sense we are not truly considering irreducible {\it representations}
of the inhomogeneous Lorentz group in this paper.  On the other hand, we are
considering what Wigner ([17], page 335) refers to as `{\it corepresentations}';
a corepresentation is just like a unitary representation only some of the
operators are allowed to be anti-unitary.  Clearly, a corepresentation is
mathematically distinct from a representation, and so it is very important not
to confuse the two things (this point is emphasized in [8]).  Technically, then,
this paper is concerned strictly with corepresentations of the inhomogeneous
Lorentz group.{\footnote{See also [20] for another discussion of these
issues.}}

The above discussion is very general and can be applied in a wide range of
situations. We now wish to specialise and
concentrate our attention on the one group which will survive in any field
theory
which incorporates relativistic covariance with discrete transformations: the
inhomogeneous
Lorentz group, $O(3, 1)$.

The best way to illustrate what we are talking about is with an explicit
example. Let
us therefore recall how the operators $C$ (charge conjugation), $P$ (parity
inversion)
and $T$ (time reversal) are represented in the particle physics
literature
[7]: 
Let ${\cal H}$ be the set of solutions of the Dirac equation on
four-dimensional
Minkowski space; then $C$, $P$, and $T$ are operators on ${\cal H}$
given by
the explicit formulae:}
\begin{eqnarray}
C: {\psi}(x, t) \rightarrow i{\gamma}^{2}{\psi}^{*}(x, t) \nonumber \\
P: {\psi}(x, t) \rightarrow {\gamma}^{0}{\psi}(-x, t) \\
T: {\psi}(x, t) \rightarrow {\gamma}^{1}{\gamma}^{3}{\psi}^{*}(x, -t) \nonumber
\end{eqnarray}
{where ${\psi}$ is any solution and $^{*}$ denotes the operation of
complex conjugation. We remind the reader (without going into details)
that a host of physical considerations goes into the choices made in equations
(2).
A number of other choices are possible, the key point being that the other
choices are
{\it mathematically inequivalent}.

Now, one of the first things we can notice about the operators $P$ and $T$
defined in
(2) is that they do not give a {\it Cliffordian} representation of the action
of
space and time inversion. That is, $P$ and $T$ do not anti-commute, 
since in fact they
commute:}
\[
PT \,\sim \, {\gamma}^{0}{\gamma}^{1}{\gamma}^{3} =
{\gamma}^{1}{\gamma}^{3}{\gamma}^{0} \,\sim \, TP
\]
{Therefore, the operators $P$ and $T$ defined in (2) correspond to a {\it
non-Cliffordian} representation of $O(3, 1)$ with non-Cliffordian {\it action}.

This situation can be contrasted with the case where the representation has
Cliffordian action. For example, a Cliffordian action can be recovered by the
following
operator assignment:}
\begin{eqnarray}
P: {\psi}(x, t) \rightarrow {\gamma}^{1}{\psi}(-x, t) \nonumber \\
\\
T: {\psi}(x, t) \rightarrow {\gamma}^{0}{\psi}(x, -t) \nonumber
\end{eqnarray}
{Clearly, the (unitary) choices in (3) anti-commute.

Of course, in each of the above examples, the underlying group structure is
identical.
More precisely, in the operator assignments made in (2), we used the group of
elements
${\gamma}^{\mu}$ satisfying $\{{\gamma}^{\mu}, {\gamma}^{\nu}\} = 2g^{\mu \nu}$
to
construct operators $P$ and $T$ whose {\it action} on ${\cal H}$ is
non-Cliffordian,
whereas in (3) we used the {\it same group} of Cliffordian elements to
construct
operators $P$ and $T$ with {\it Cliffordian} action. It is absolutely essential
that
we make this distinction between the different actions on a Hilbert space which
can
be constructed from a given group, and genuinely {\it different groups}. This
is
because we are sympathetic to the philosophy of Wigner [4] who put forward the
idea
that the irreducible (co)representations of whatever group of symmetries is
present in nature
should form the basis for any theory of elementary particles. Indeed, Wigner
completely classified the set of irreducible corepresentations of the
inhomogeneous
Lorentz group, $O(3, 1)$, on the Hilbert space of solutions to the Dirac
equation (1)
with $m \,\neq \, 0$. He showed that once one `fixes' the sign of the square of
parity
inversion $P^{2}$ (fixing this sign corresponds to choosing a signature for
spacetime,
basically) then there are four {\it inequivalent} (non-isomorphic) cases. The
first
case is the standard particle physics choice made in (2) above. In the
remaining three
cases, there is a phenomenon known as `parity doubling', which can be described
as
follows.

To begin with, there are simply not enough choices possible, when the dimension
of the
corepresentation is 4, to realise all of the irreducible corepresentations. That is
to
say, if we stick with only using $4 \times 4$ matrices to write $P$ and $T$ as
linear
operators on ${\cal H}$, then we can really only use combinations of the
${\gamma}^{\mu}$s and so we are stuck with the standard Cliffordian {\it group}
which
we used in examples (2) and (3) above. We therefore need to somehow increase
the
dimension of our corepresentation, and in fact this is exactly what Wigner did
when he
showed how to obtain the remaining three corepresentations by {\it doubling} the
dimension.

Explicitly, what one first does is write down the `doubled' gamma matrices,
${\Gamma}^{\mu}$ (the `big' gammas) as follows:}
\begin{equation}
{\Gamma}^{\mu} = \left( \begin{array}{cc}
{\gamma}^{\mu} &0 \\
0 &{\gamma}^{\mu}
\end{array} \right)
\end{equation}

{The `doubled' Dirac equation then becomes}
\begin{equation}
(i{\Gamma}^{\mu}\, {\partial}_{\mu} - m)\,{\psi} = 0
\end{equation}
{Thus, solutions to (5) are now {\it eight} component `pinor' fields.
Intuitively, one
can now think of the extra degrees of freedom in the solutions of (5) as
corresponding to the assignment of `parity'.

We can now obtain the non-standard irreducible corepresentations of $O(3, 1)$ by
representing $P$ and $T$ on the set of solutions, ${\cal H}_{D}$ (the `doubled'
Hilbert space), to (5). Of course, this might seem confusing since although
the
${\Gamma}^{\mu}$s are eight component matrices, they still satisfy}
\[
\{{\Gamma}^{\mu}, {\Gamma}^{\nu}\} = 2g^{\mu \nu}
\]
{The point is, we are {\it no longer bound} to only use combinations of the
${\Gamma}^{\mu}$s to construct our corepresentations. The only thing [4] which
distinguishes the different irreducible corepresentations (once we have fixed the
signature) is the {\it sign of the squares of the operators representing $T$
and
$PT$}. Let us fix the signature to be (for now) $(- + + +)$. Then the sign of
parity
inversion squared is fixed (in all the corepresentations) to be}
\[
P^{2} = -{\mbox{Id}}
\]

{Thus, in the `standard' case presented above (which we shall denote {\it Case
I})
$P^{2} = {\gamma}^{0}\,{\gamma}^{0} = -{\Bbb I}$, $T^{2} =
{\gamma}^{1}\,{\gamma}^{3}\,
{\gamma}^{1}\,{\gamma}^{3} = -{\Bbb I}$, $(PT)^{2} =
{\gamma}^{0}\,{\gamma}^{1}\,
{\gamma}^{3}\,{\gamma}^{0}\,{\gamma}^{1}\,{\gamma}^{3} = {\Bbb I}$ where ${\Bbb
I} =
\mbox{Id}$ is the identity matrix. The other three cases can therefore be
presented as
follows.}\\
\vspace*{0.2cm}

{\noindent {\it Case II:} ~Here, we seek operators $P$, $T$, and $PT$ on
the
space of solutions ${\cal H_{D}}$ to (5) such that $P^{2} = -{\Bbb I}$, $T^{2}
=
-{\Bbb I}$, and $(PT)^{2} = -{\Bbb I}$. Such a corepresentation is given by the
following assignments:}
\begin{eqnarray}
P: {\psi}(x, t) \rightarrow \left( \begin{array}{cc}
{\gamma}^{0} &0 \\
0 &-{\gamma}^{0}
\end{array} \right) \, {\psi}(-x, t) \nonumber \\
\\
T: {\psi}(x, t) \rightarrow \left( \begin{array}{cc}
0 &{\gamma}^{1}\,{\gamma}^{3} \\
{\gamma}^{1}\,{\gamma}^{3} &0
\end{array} \right) \, {\psi}^{*}(x, -t) \nonumber
\end{eqnarray}
\vspace*{0.2cm}

{\noindent {\it Case III:} ~Here, we seek operators $P$ and $T$  such that
$P^{2} = -{\Bbb I}$, $T^{2} =
+{\Bbb I}$, and $(PT)^{2} = +{\Bbb I}$. Such a corepresentation is given by the
following assignments:}
\begin{eqnarray}
P: {\psi}(x, t) \rightarrow \left( \begin{array}{cc}
{\gamma}^{0} &0 \\
0 &-{\gamma}^{0}
\end{array} \right) \, {\psi}(-x, t) \nonumber \\
\\
T: {\psi}(x, t) \rightarrow \left( \begin{array}{cc}
0 &{\gamma}^{1}\,{\gamma}^{3} \\
-{\gamma}^{1}\,{\gamma}^{3} &0
\end{array} \right) \, {\psi}^{*}(x, -t) \nonumber
\end{eqnarray}
\vspace*{0.2cm}

{\noindent {\it Case IV:} ~Finally, in this case we seek operators $P$ and $T$
for
which $P^{2} = -{\Bbb I}$, $T^{2} =
+{\Bbb I}$, and $(PT)^{2} = -{\Bbb I}$. This is accomplished by the
following definitions:}
\begin{eqnarray}
P: {\psi}(x, t) \rightarrow \left( \begin{array}{cc}
{\gamma}^{0} &0 \\
0 &{\gamma}^{0}
\end{array} \right) \, {\psi}(-x, t) \nonumber \\
\\
T: {\psi}(x, t) \rightarrow \left( \begin{array}{cc}
0 &{\gamma}^{1}\,{\gamma}^{3} \\
-{\gamma}^{1}\,{\gamma}^{3} &0
\end{array} \right) \, {\psi}^{*}(x, -t) \nonumber
\end{eqnarray}

{Of course, if we change the signature (or just the sign of $P^{2}$) then we
again
obtain four inequivalent corepresentations. These eight different ways of writing
the
operations $P$ and $T$ thus correspond to eight different non-isomorphic
groups. These
groups are called the {\it pin groups}, and it is time we turned our attention
to
formally defining them.

To this end, recall that generally we do physics on {\it spacetimes}, $M$, which
may
not necessarily be orientable. What this means is that the tangent bundle,
${\tau}_{M}$,
can at most be reduced to an $O(p, q)$ bundle. When the metric, $g_{ab}$, has
signature $(- + + +)$ then the structure group will be $O(3, 1)$. When the
metric has
signature $(+ - - -)$ then the structure group will be $O(1, 3)$ (actually,
$O(3, 1)
{}~{\simeq}~ O(1, 3)$, but as we shall see it is necessary to keep the
distinction when
we pass to the double covers). Since ${\pi}_{1}(O_{0}(3, 1) ~{\simeq}~
{\pi}_{1}(O_{0}(1, 3)) ~{\simeq}~ {\Bbb Z_{2}}$, we are interested in finding
all groups
which are double covers of $O(3, 1)$ and $O(1, 3)$. There are
{\it eight} distinct such double covers [6] of $O(p, q)$. Following D\c{a}browski,
we will write these covers as}
\[
{h^{a, b, c}:~ {\mbox{Pin}}^{a, b, c}(p, q) ~{\longrightarrow}~ O(p, q)}
\]
{with $a, b, c ~{\in}~ {\{}+, -{\}}$. The signs of $a, b$, and $c$ can be
interpreted
in the following way:

Recall, first, that $O(p, q)$ is not path connected; there are four components,
given
by the identity connected component, $O_{0}(p, q)$, and the three components
corresponding to parity reversal $P$, time reversal $T$, and the combination of
these
two, $PT$ (i.e., $O(p, q)$ decomposes into a semidirect product{\footnote{That
is,
$O(p, q)$ is the disjoint union $O(p, q)$ $=$ $(O_{0}(p, q))$ $~{\cup}~$
$P(O_{0}(p,
q))$ $~{\cup}~$ $T(O_{0}(p, q))$ $~{\cup}~$ $PT(O_{0}(p, q))$, and the four
element
group ${\{}1, P, T, PT{\}}$ is isomorphic to ${\Bbb Z_{2}} ~{\times}~ {\Bbb
Z_{2}}$.}}, $O(p, q) ~{\simeq}~ O_{0}(p, q) ~{\odot}~ ({\Bbb Z_{2}} ~{\times}~
{\Bbb Z_{2}})$). The signs of $a, b$, and $c$ then correspond to the signs of
the
squares of the elements in ${\mbox{Pin}}^{a, b, c}(p, q)$ which cover space
reflection,
$R_{S}$, time reversal, $R_{T}$ and a combination of the two respectively.
That is, in this paper we adopt precisely the
following
convention:}
\[
\begin{array}{c}
P^{2} = a \\
T^{2} = b \\
(PT)^{2} = c
\end{array}
\]
{We note that this convention differs markedly from D\c{a}browski, who takes}
\[
\begin{array}{c}
P^{2} = -a \\
T^{2} = b \\
(PT)^{2} = -c
\end{array}
\]
{(In our notation, the obstruction theory is more transparent, although there
are
other reasons for adopting D\c{a}browski's notation).

With this in mind we can, following D\c{a}browski [6], write out the explicit form
of the
groups ${\mbox{Pin}}^{a, b, c}(p, q)$; they are given by the semidirect
product}
\[
{{\mbox{Pin}}^{a, b, c}(p, q) ~{\simeq}~ {\frac{({\mbox{Spin}}_{0}(p, q)
{}~{\odot}~
C^{a, b, c})}{{\Bbb Z_{2}}}}}
\]
{where the $C^{a, b, c}$ are the four double coverings of ${\Bbb Z_{2}}
{}~{\times}~
{\Bbb Z_{2}}$; i.e., $C^{a, b, c}$ are the groups ${\Bbb Z_{2}} ~{\times}~
{\Bbb
Z_{2}} ~{\times}~ {\Bbb Z_{2}}$ (when $a = b = c = +$), $D_{4}$ (dihedral
group, when
there are two plusses and one minus in the triple $a, b, c$), ${\Bbb Z_{2}}
{}~{\times}~
{\Bbb Z_{4}}$ (when there are two minuses and one plus in $a, b, c$), and
$Q_{4}$
(quaternions, when $a = b = c = -$). 

Clearly, the different pin groups correspond to the different ways of defining
the operators $P$ and $T$.  We shall therefore say that the 
different pin groups determine different ${\it types}$ of fermions.  Our goal now 
is to explore the extent to which fermion type defines a superselection rule, i.e.,
is it possible to form a coherent superposition of fermions of different type?}\\
\vspace*{0.6cm}

{\noindent \bf 3. Fermion Type and Superselection}\\

{Traditionally, people have assumed that fermion type determines a superselection
rule, i.e., that it is impossible to form a linear combination of fermionic states
of differing type.  This prejudice is based primarily on the fact that the
different pin groups form all of the ${\it irreducible}$ representations of the
inhomogeneous Lorentz group.  Thus, any attempt to mix fermions of differing
type will require passing to a manifestly non-irreducible representation.

In order to rigorously see why fermion type determines a superselection rule,
it would be nice if we could write down an equation similar to the one used
in the introduction to show that the observable $(-1)^{F}$ yields
superselection. To do this, let ${\cal H}^{a, b, c}$ denote the Hilbert space
for a particle of type $(a, b, c)$ acted on by $\mbox{Pin}^{a, b, c}(3, 1)$.
Let $P_{(a, b, c)}$ and $T_{(a, b,, c)}$ denote the operations of parity and time
reversal in $\mbox{Pin}^{a, b, c}(3, 1)$. Consider two fermions of distinct
type, ${\psi}_{+} \,\in\, {\cal H}^{+, b, c}$ and ${\psi}_{-} \,\in\, 
{\cal H}^{-, b, c}$. We want to know if it makes sense to form the linear
combination $\alpha\psi_{+} \,+\, \beta\psi_{-}$. Na{\" \i}vely then, we want
to consider an expression of the form}
\[
P^{2}(\alpha\psi_{+} \,+\, \beta\psi_{-})
\]
{However, an obvious problem which presents itself is: Which `$P$' do we
choose? Clearly, it makes no sense mathematically to have either $P = 
P_{(+, b, c)}$ {\it or} $P = P_{(-, b, c)}$. It {\it does} make sense to write}
\[
P = \left( \begin{array}{cc}
P_{(+, b, c)} & 0 \\
0 & P_{(-, b, c)}
\end{array}
\right)
\]
{and to think of $\psi_{+}$ and $\psi_{-}$ as two `states' of a `larger'
particle $\Phi$:}
\[
\Phi = \left( \begin{array}{cc}
\alpha \psi_{+} \\
\beta \psi_{-}
\end{array}
\right)
\]

{In fact, not only does this construction make sense, it is mathematically
justified; to see this, consider the following thought experiment:}

{Suppose we are given a system consisting of two particles of type $(a, b,
c)$ and two particles of different type $(a^{\prime}, b^{\prime},
c^{\prime})$. Then the appropriate Hilbert space for such a system is}
\begin{equation}
\left({\cal H}^{a, b, c} \,\textcircled{\scshape A}\, {\cal H}^{a, b, c}\right) \,\oplus\, 
\left({\cal H}^{a, b, c} \,\otimes\, {\cal H}^{a^{\prime}, b^{\prime}, 
c^{\prime}}\right) \,\oplus\, \left({\cal H}^{a^{\prime}, b^{\prime}, 
c^{\prime}} \,\textcircled{\scshape A}\, {\cal H}^{a^{\prime}, b^{\prime}, c^{\prime}}\right)
\end{equation}
{where \textcircled{\scshape A} denotes antisymmetric product, $\oplus$ denotes direct sum and
$\otimes$ denotes tensor product. (In other words, the pair of $(a, b, c)$
particles and the pair of $(a^{\prime}, b^{\prime}, c^{\prime})$ particles
each satisfy Pauli exclusion since they are each pairs of identical
particles). The beautiful thing
is that the Hilbert space in Equation (9) is actually {\it isomorphic} to the
Hilbert space}
\[
\left({\cal H}^{a, b, c} \,\oplus\, {\cal H}^{a^{\prime}, b^{\prime},
c^{\prime}}\right) \,\textcircled{\scshape A}\, \left({\cal H}^{a, b, c} \,\oplus\, {\cal H}^{a^{\prime},
b^{\prime}, c^{\prime}}\right)
\]
{In other words, it is mathematically equivalent to think of the four
particle system as a {\it two} particle system consisting of two fermions,
each living in the Hilbert space $\left({\cal H}^{a, b, c} \,\oplus\, 
{\cal H}^{a^{\prime}, b^{\prime}, c^{\prime}}\right)$. We shall refer to
fermions which live in such direct sum Hilbert spaces as `mixed' fermions or
{\it meta}-fermions. In the language of group theory these objects are {\it pinor
supermultiplets}, since each `state' of the multiplet is an object corresponding
to a distinct pin group; we emphasize that this use of the word `supermultiplet'
has nothing to do with supersymmetry, i.e., we are using the terminology 
of Chapter 18 of [19].
Thus, by passing to the space of mixed fermions we can
considerably simplify the mathematical structure of a problem (although we
are still dealing with the same amount of information). Of course, in general
there will be eight (not just two) types of fermion present; suppose that the
total number of fermions (of whatever type) is $N$. Then the generalisation
of the above Hilbert space isomorphism implies that we can always think of
such a system as consisting of $N$ identical particles, each living in the
Hilbert space}
\[
\stackrel{\oplus}{\scriptsize \mbox{$(a, b, c) \in \{\pm\}$}} 
{\cal H}^{a, b, c}
\]
{In other words, the general Hilbert space for fermions is}
\begin{equation}
\underbrace{\left( \stackrel{\oplus}{\scriptsize \mbox{$(a, b, c) \in \{\pm\}$}} {\cal H}^{a, b, c} \right)
\textcircled{\scshape A} \left( \stackrel{\oplus}{\scriptsize \mbox{$(a, b, c) \in \{\pm\}$}} {\cal H}^{a, b, c} 
\right) \textcircled{\scshape A} \dots \textcircled{\scshape A} \left( \stackrel{\oplus}{\scriptsize \mbox{$(a, b, c) \in 
\{\pm\}$}} {\cal H}^{a, b, c} \right)} \atop {N ~\mbox{times}}
\end{equation}

{Clearly, this proposal is very similar to Heisenberg's old suggestion [9]
that we should think of the proton $p$ and the neutron $n$ as two `states' of
a single particle, the nucleon $N$:}
\[
N = \left( \begin{array}{c}
p \\
n
\end{array}
\right)
\]
{Of course, Heisenberg took things further, introducing the abstract `isospin
space', defining the proton to be isospin up and the neutron to be isospin
down, and proposing that strong interaction physics is invariant under
rotations in isospin space. In other words, in terms of group theory, he
asserted that strong interactions are invariant under the action of an
internal symmetry $SU(2)$, and that nucleons determine a two-dimensional
representation (i.e., they are isospin $\frac{1}{2}$). This proposal, which
was motivated by the simple fact that strong interactions do not distinguish
between protons and neutrons, had far-reaching consequences.

To our knowledge, {\it none} of the four forces distinguish between fermions
because of type; indeed, the only `physical' effect of fermion type known to us
(we will discuss this in more detail later) is the fact [5] that some types
of fermions {\it do not} have CP-violating electric dipole moments whereas
other types do. Given this, it is tempting to {\it conjecture} that any
physics involving the mixed fermion supermultiplet which we constructed above
is invariant under the maximal internal symmetry group $U(8)$. If this were
true, then the supermultiplet would form a {\it fundamental}
(eight-dimensional) representation of $U(8)$. On the other hand, it may be
that some physical processes break the symmetry down to some $(S)U(n)$, $n <
8$. We simply cannot tell since we have no real experimental data which
determines fermion type and, more seriously, we do not even know if there is
more than one type of fermion in the universe. Nevertheless, it is amusing to
take these abstract group-theoretic conjectures seriously and see if they
might lead us to any real physics; this avenue of research is currently being
actively investigated. We will have more to say about the possible
experimental consequences of this proposal that fermions live in
eight-dimensional supermultiplets later.

We conclude this section with a sketch of the structure which we have
proposed:

A fermion $\Psi$ generically lives in a direct sum of Hilbert spaces ${\cal
H}_{a, b, c}$, where each ${\cal H}^{a, b, c}$ is acted on by a
representation of the relevant pin group $\mbox{Pin}^{a, b, c}(p, q)$.
Explicitly, $\Psi$ looks like this:}
\begin{equation}
\Psi = \left( \begin{array}{c}
\Psi_{+\,+\,+} \\
\Psi_{+\,-\,+} \\
\Psi_{+\,-\,-} \\
\vspace*{-2.6mm}

\vdots \\
\vdots \\
\Psi_{-\,-\,-}
\end{array}
\right)
\end{equation}
{We emphasize that this is only a {\it proposal}.  It may well be the case
that every electron (for example) in the universe 
lives in a Hilbert space acted on by just
one of the pin groups.  If this turns out to be the case, then the hypothesis
of pinor multiplets is a needless complication.  On the other hand, it may well
be the case that some electrons are of type $(+, +, +)$, whereas other electrons
are of type $(+, -, +)$, and so forth.  If this turns out to be the case,
then we have to assign extra internal quantum numbers (namely the three signs
for $a$, $b$ and $c$) to any electron in order to completely classify the state.

$\Psi$ is acted upon by a `total' parity, or metaparity operator, $P$, which
also is a direct sum of the individual parity operators $P_{(a, b, c)}$
coming from each $\mbox{Pin}^{a, b, c}(p, q)$, i.e.,}

\begin{equation}
P = \left( \begin{array}{cc}
\begin{array}{ccc}
P_{(+\,+\,+)} & & \\
 &P_{(+\,-\,+)} & \\
 & &P_{(+\,-\,-)}
\end{array} & \mbox{\Huge 0} \\
\hspace*{-19mm} \mbox{\Huge 0} &\begin{array}{ccc}
\vspace*{-3.5mm}

\hspace*{-0.5mm}\ddots & & \\
 &\hspace*{-3.1mm} \ddots & \\
 & &P_{(-\,-\,-)}
\end{array}
\end{array} \right)
\end{equation}
{Similarly, there is a total time inversion, $T$, which looks like}
\begin{equation}
T = \left( \begin{array}{cc}
\begin{array}{ccc}
T_{(+\,+\,+)} & & \\
 &T_{(+\,-\,+)} & \\
 & &T_{(+\,-\,-)}
\end{array} & \mbox{\Huge 0} \\
\hspace*{-19mm} \mbox{\Huge 0} &\begin{array}{ccc}
\vspace*{-3.5mm}

\hspace*{-0.5mm}\ddots & & \\
 &\hspace*{-3.2mm} \ddots & \\
 & &T_{(-\,-\,-)}
\end{array}
\end{array} \right)
\end{equation}
{and similarly for $PT$.

In the absence of any interaction, propagation in each Hilbert space is given
by the ordinary Dirac equation. This would seem to justify the supposition
that the different ${\cal H}^{a, b, c}$ determine superselection sectors for
fermions, i.e., that different fermion types cannot interfere. However, it is
likely that an argument similar to the one given by Aharonov and Susskind [3]
can be constructed to explicitly show how to prepare states which are
coherent superpositions of fermions of differing types. Recall that
Aharonov and Susskind presented a thought experiment, which could be {\it
performed} in principle, in which they showed how to prepare a state which is
a coherent superposition of a proton and a neutron:}
\[
\alpha p \,+\, \beta n
\]
{Since the proton and neutron components of this nucleon can interfere, this
amounts to a violation of the charge superselection rule. It is likely that a
similar thought experiment can be conceived for fermion type; it is probably
just a question of understanding how to distinguish (in the lab) and isolate,
fermions of differing types.

With this in mind, we now turn to a discussion of what observable
consequences (if any) follow from our proposal that real fermions actually
belong to these eight-dimensional pinor supermultiplets.}\\

\newpage
{\noindent \bf 4. CP-Invariance and Electric Dipole Moments}\\

{A great deal of experimental evidence has been amassed which establishes
very strong bounds on the electric dipole moments of various elementary
particles [11], [12]. In particular, it has been shown that the electric
dipole moment (e.d.m.) of the electron (denoted $d_{e}$) satisfies [11]}
\begin{equation}
d_{e} < (-0.3 \pm 0.8) \,\times\, 10^{-26} ~\mbox{ecm}
\end{equation}
{and that the e.d.m. of the neutron (denoted $d_{n}$) satisfies}
\begin{equation}
d_{n} < 11 \,\times\, 10^{-26} ~\mbox{ecm} ~.
\end{equation}
{Clearly, these bounds imply that the e.d.m.'s of these particles are {\it
extremely} small, even smaller than the particles themselves ($\sim
10^{-13}$ cm for the neutron $n$). It is very important that we know the {\it
precise} value of $d_{n}$ since it is related to other quantities which arise
naturally in the standard model $(SM)$.

For example, the $SU(3) \,\times\, SU(2) \,\times\, U(1)$ gauge sector of
$SM$ has a non-trivial vacuum structure [13]. This vacuum structure gives
rise to phases (or `$\theta$-vacua' [14]) which imply the existence of
CP-violating effective interaction terms, which involve the non-Abelian gauge
fields:}
\[
{\cal L}_{\mbox{\scriptsize eff}} \,\simeq\, \theta_{s}\,
\frac{\alpha_{s}}{8\pi}\, F_{a}^{\mu \nu}\, {\tilde F}_{a \mu \nu} \,+\,
\theta_{w}\, \frac{\alpha_{w}}{8\pi}\, W_{a}^{\mu \nu}\, W_{a \mu \nu}
\]
{Since the electroweak theory is chiral, we can always rotate the {\it weak}
vacuum angle, $\theta_{w}$, to zero. However, the {\it strong} vacuum angle
$\theta_{s}$ is more complicated; one has to perform chiral rotations that
leave the quark mass matrices diagonal. This means that $\theta_{s}$ receives
corrections from the weak sector:}
\[
{\bar \theta} = \theta_{s} \,+\, \arg (\det M)
\]
{where $M$ is the quark mass matrix. In other words, the {\it physical}
CP-violating interaction is}
\[
{\cal L}_{\mbox{\scriptsize CP}} = {\bar \theta}\, \frac{\alpha_{s}}{8\pi}\,
 F_{a}^{\mu \nu}\,
{\tilde F}_{a \mu \nu}
\]
{Interestingly [16], the existence of such an interaction in $SM$ contributes
substantially to the neutron e.d.m., $d_{n}$. In fact,}
\begin{equation}
d_{n} \approx 8.2~{\times}~ 10^{-16} {\bar \theta} ~\mbox{ecm}
\end{equation}
{Actually, this estimate is based on a calculation in QCD, and it {\it assumes}
that the e.d.m. of the neutron is CP-violating and of course that the neutron 
is a system made up of three quarks. 
Perhaps the truly interesting thing to do here is to try and 
repeat the calculation of [16] while allowing for the quarks themselves
to be particles of differing type.  In this paper we are being more
simple minded about things and regarding the neutron itself as an elementary
particle.  At any rate, given the above
bound equation (17) on $d_{n}$, we see that ${\bar \theta}$ must satisfy [11]}
\begin{equation}
{\bar \theta} \,\leq\, 10^{-9} \,-\, 10^{-10}
\end{equation}
{Finding an explanation for this phenomenon is known as the {\it strong CP
problem}.

While we do not solve the strong CP problem here, we {\it do} present proof
that fermions of differing type possess e.d.m.'s which break {\it differing}
combinations of C, P, or T. More precisely, we show that by choosing
different types we can construct fermions with e.d.m.'s  which are {\it not}
CP-violating, but which may be (for example) C-violating as well as
P-violating (hence T-non-violating by the CPT theorem). In order to
understand this construction, we need to recall the recent work of Giesen [5].

In [5] Giesen studied the behaviour of the e.d.m.'s of particles of differing
types under the action of discrete space-time symmetries. What he found is
that while the `standard' four-component fermions (in the chiral
representation) of type $(+, -, -)$ possess e.d.m.'s which are both P and T
violating (and hence CP violating), the `non-standard' eight-component
fermions of type $(+, +, -)$ (the $a = P^{2} = +1$ analogue of Case III,
equation (7) above) possess e.d.m.'s which are {\it neither} P nor T
violating (and hence do {\it not} violate CP).

In order to make everything explicit, we write out the actions of C, P, and T
for fermion types with $P^{2} = +1$ in the below table (this table is the $a
= P^{2} = +1$ analogue of equations (2), (6), (7), (8) above).  Throughout
this section we are working in the `chiral' representation, i.e.,}
\[
{\gamma}^{0} = \left( \begin{array}{cc}
0 &{\Bbb I} \\
{\Bbb I} &0
\end{array} \right)
\]
and 
\[
{\gamma}^{i} = \left( \begin{array}{cc}
0 &{\sigma}_{i} \\
-{\sigma}_{i} &0
\end{array} \right)
\]

where the ${\sigma}_{i}$ are ordinary Pauli matrices.  The actions of the
different discrete transformations are then given as shown:

\begin{center}
\begin{tabular}{|c|l|}
\hline
 & \\
Fermion Type &\hspace*{3.5cm} Actions of C, P, and T \\
 & \\
\hline
 & \\
$(+, -, -)$ &$\left. \begin{array}{l}
C: \psi(x, t) \longrightarrow i \gamma^{2} \psi^{*}(x, t) \\
P: \psi(x, t) \longrightarrow \gamma^{0} \psi (-x, t) \\
T: \psi(x, t) \longrightarrow \gamma^{1} \gamma^{3} \psi^{*} (x, -t) 
\end{array} \right\}$ $\begin{array}{l}
\mbox{Four-component} \\
\mbox{corepresentation}
\end{array}$ \\
 & \\
\hline
 & \\
$(+, -, +)$ &$\left. \begin{array}{l} 
C: \psi(x, t) \longrightarrow i \left( \begin{array}{cc}
\gamma^{2} &0 \\
0 &- \gamma^{2}
\end{array} \right)
\psi^{*} (x, t) \\
P: \psi(x, t) \longrightarrow \left( \begin{array}{cc}
\gamma^{0} &0 \\
0 & -\gamma^{0}
\end{array} \right)
\psi (-x, t) \\
T: \psi(x, t) \longrightarrow \left( \begin{array}{cc}
0 &\gamma^{1}\gamma^{3} \\
\gamma^{1}\gamma^{3} &0
\end{array} \right)
\psi^{*} (x, -t) 
\end{array} \hspace*{3mm} \right\}$ $\begin{array}{l}
\mbox{Doubled} \\
\mbox{corepresentation} \\
\mbox{(eight-component)}
\end{array}$ \\
 & \\
\hline
 & \\
$(+, +, -)$ &$\left. \begin{array}{l} 
C: \psi(x, t) \longrightarrow i \left( \begin{array}{cc}
\gamma^{2} &0 \\
0 &\gamma^{2}
\end{array} \right)
\psi^{*} (x, t) \\
P: \psi(x, t) \longrightarrow \left( \begin{array}{cc}
\gamma^{0} &0 \\
0 & -\gamma^{0}
\end{array} \right)
\psi (-x, t) \\
T: \psi(x, t) \longrightarrow \left( \begin{array}{cc}
0 &\gamma^{1}\gamma^{3} \\
- \gamma^{1}\gamma^{3} &0
\end{array} \right)
\psi^{*} (x, -t) 
\end{array} \right\}$ $\begin{array}{l}
\mbox{Doubled} \\
\mbox{corepresentation} \\
\mbox{(eight-component)}
\end{array}$ \\
 & \\
\hline
 & \\
$(+, +, +)$ &$\left. \begin{array}{l} 
C: \psi(x, t) \longrightarrow i \left( \begin{array}{cc}
\gamma^{2} &0 \\
0 &- \gamma^{2}
\end{array} \right)
\psi^{*} (x, t) \\
P: \psi(x, t) \longrightarrow \left( \begin{array}{cc}
\gamma^{0} &0 \\
0 &\gamma^{0}
\end{array} \right)
\psi (-x, t) \\
T: \psi(x, t) \longrightarrow \left( \begin{array}{cc}
0 &\gamma^{1}\gamma^{3} \\
- \gamma^{1}\gamma^{3} &0
\end{array} \right)
\psi^{*} (x, -t) 
\end{array} \right\}$ $\begin{array}{l}
\mbox{Doubled} \\
\mbox{corepresentation} \\
\mbox{(eight-component)}
\end{array}$ \\
 & \\
\hline
\end{tabular}
\end{center}
\begin{center}
{\bf Table 1}
\end{center}
\newpage

{In the above table, $\{ \gamma^{\mu}, \gamma^{\nu} \} =  2g^{\mu \nu}$,
with $g^{\mu \nu}$ of signature $(+, -, -, -)$ (so that $P^{2} = +1$
everywhere).

To see how the e.d.m.'s of particles of different type transform, we follow
[5] and write the Dirac equation for a four-component fermion $\psi$ with
dipole moment strength $d$ coupled to an external electromagnetic field
$A_{\mu}$ (with field strength $F_{\mu \nu}$) as follows:}
\begin{equation}
\left( \gamma^{\mu} (i\partial_{\mu}  \,+\, eA_{\mu}) \,-\, d \gamma^{\mu}
\gamma^{\nu} \gamma_{5} F_{\mu \nu} \,-\, m \right) \psi = 0
\end{equation}
{where $\gamma_{5} = i\gamma^{0}\gamma^{1}\gamma^{2}\gamma^{3}$. The extra
term in this otherwise minimally coupled Dirac equation comes from the
addition of a gauge invariant, covariant effective Lagrangian term}
\[
{\cal L}_{\mbox{\scriptsize eff}} = - d\psi^{\dagger}
\gamma^{\mu}\gamma^{\nu} \gamma_{5} F_{\mu \nu} \psi
\]
{The non-relativistic limit of this coupling is the usual ${\hat \sigma} 
\cdot E$ type interaction, where}
\begin{eqnarray*}
{\hat \sigma} = \left( {\hat \sigma}_{1}, {\hat \sigma}_{2}, {\hat
\sigma}_{3} \right) \\
 \\
{\hat \sigma}_{i} = \left( \begin{array}{cc}
\sigma_{i} &0 \\
0 &\sigma_{i}
\end{array} \right)
\end{eqnarray*}
{and $\sigma_{i}$ are the Pauli matrices.

Let $\psi_{P} = P\psi$ denote the parity inversion of $\psi$, and $\psi_{T} =
T\psi$ the time inversion of $\psi$. Then it is a standard result [5] that
$\psi_{P}$ is {\it not} a solution of the parity reflection of equation (18),
and similarly $\psi_{T}$ is {\it not} a solution of the time reflection
of equation (18). Thus,
reflected solutions do not solve the reflected equation; we therefore say
that solutions of (18) violate P and T symmetry. This is an old result, which
holds for the e.d.m.'s of all four-component fermions of type $(+, -, -)$.

However, things change considerably when we write down the equation
describing a dipole moment for a {\it non-standard} eight-component fermion
[5]:}
\begin{equation}
\left( \Gamma^{\mu} (i\partial_{\mu} \,+\, eA_{\mu}) \,-\, d\left(
\begin{array}{cc}
0 &\gamma^{\mu}\gamma^{\nu}\gamma_{5} \\
\gamma^{\mu}\gamma^{\nu}\gamma_{5} &0
\end{array} \right)
F_{\mu \nu} \,-\, m\right) \psi = 0
\end{equation}
{where $\Gamma^{\mu} = \left( \begin{array}{cc}
\gamma^{\mu} &0 \\
0 &\gamma^{\nu}
\end{array} \right)$ are the doubled gamma matrices. Equation (19) arises by
adding the effective Lagrangian term}
\[
{\cal L}_{\mbox{\scriptsize eff}} = -d \psi^{\dagger} \left(
\begin{array}{cc}
0 &\gamma^{\mu}\gamma^{\nu}\gamma_{5} \\
\gamma^{\mu}\gamma^{\nu}\gamma_{5} &0
\end{array} \right)
F_{\mu \nu} \psi
\]
{In the non-relativistic limit, this coupling takes the form}
\[
\left( \begin{array}{cc}
0 &{\hat \sigma} \\
{\hat \sigma} &0
\end{array} \right)
\cdot E
\]
{with ${\hat \sigma}$ as above.

For fermions of type $(+, +, -)$ Giesen showed [5] that the reflected
solutions $\psi_{P}$ and $\psi_{T}$ {\it are} solutions of the P and T
inversions of equation (19). Thus, all fermions of type $(+, +, -)$ possess
e.d.m.'s which are {\it not} CP-violating.

A natural question, then, is to determine how the e.d.m's of {\it other}
types of particles transform under the action of C, P. and T. It is not too
hard to work out; the results are displayed in the below table.}
\vspace*{0.3cm}

\begin{center}
\begin{tabular}{|c|c|c|c|c|}
\hline
 &e.d.m. &e.d.m. &e.d.m. &e.d.m. \\
Fermion &violates &violates &violates &violates \\
Type &C? &P? &T? &CP? \\
\hline
 & & & & \\
$(+, -, -)$ &NO &YES &YES &YES \\
 & & & & \\
\hline
 & & & & \\
$(+, -, +)$ &YES &NO &YES &YES \\
 & & & & \\
\hline
 & & & & \\
$(+, +, -)$ &NO &NO &NO &NO \\
 & & & & \\
\hline
 & & & & \\
$(+, +, +)$ &YES &YES &NO &NO \\
 & & & & \\
\hline
\end{tabular}
\end{center}
\begin{center}
{\bf Table 2}
\end{center}
\vspace*{0.6cm}

{Clearly, what Table 2 provides us with is a way of in principle determining 
the type of an elementary particle.  For suppose that you are given any elementary
particle `$x$' with non-vanishing e.d.m. `$d$'.  Then you can determine the 
type of $x$ (up to the sign of $P^2$) simply by determining which combination
of $C$, $P$ and $T$ $d$ violates (there will be will be a table identical to 
Table 2 for the quartet of particles with $P^2 = -1$).  To our knowledge, this
is the first `in principle' performable test for determining the type of a 
fermion (but see [8] for further discussion of these points).  
The only other example where different fermion types yield different
observables was presented in [10], where it was shown that the vacuum 
expectation value of the fermionic current on a Klein bottle will depend
crucially upon which pin structure you use to construct the fermions.
Actually, we have no problem with this example since as far as we are concerned
if one accepts the path integral prescription for quantum gravity then a sum
over histories means a sum over {\it all} topologies, including non-orientable
manifolds.  Unfortunately, many people still have an aversion to
the concept of non-orientable spacetime foam.  The Giesen construction is 
therefore better for determining fermion type since it involves nothing more
than quantum mechanics on flat spacetime.
\vspace*{0.5cm}

{\noindent \bf 5. Conclusion: Does M-theory select a unique pin structure?}\\

We have attempted to determine the logical consequences of the proposal that
elementary particles should be classified according to how they behave under
the action of the full inhomogeneous Lorentz group.  We have argued that if 
more than one `type' of particle actually occurs in nature, then it is simplest
to arrange the different types into `mixed' particles, or multiplets.
We have also examined and extended Giesen's work on the nature of the electric
dipole moments of elementary particles of differing types.  We have shown that
the type of any fermion $x$ with non-vanishing e.d.m. can be determined once one
knows which combination of $C$, $P$, or $T$ invariance $x$ violates
when it interacts with
an external electromagnetic field.  We have argued that the next logical thing
to do is to repeat the calculation of [16] for the neutron e.d.m., allowing
the quarks to be of any type.  

Of course, it 
is not hard to see that most of the observed elementary particles
can only come in one type.  For example, suppose that there existed two
types of electron, a `plus' type and a `minus' type.  The Pauli exclusion
principle would allow you to place a plus electron and a minus electron
in the same state.  Obviously, this would seriously mess up most of known
chemistry unless the electromagnetic interaction coupled only to
one type, and the other type was decoupled from known matter!
Thus, it would seem that nature has selected a particular pin structure
for the description of elementary particles.  From a four-dimensional
point of view, it is unclear why or how nature makes such a selection.

In the search for some mechanism which could give rise to such a selection
rule, it is natural to appeal to some fundamental theory which might
be valid at arbitrarily high energies.  At the present time, our best
hope for such a `theory of everything' is the body of knowledge commonly
referred to as `M-theory'.  While we still aren't really sure about what
M-theory actually is, or what it describes in general, we are sure that
the low energy limit is $D = 11$, $N = 1$ supergravity theory.

Now, $D = 11$, $N = 1$ supergravity
is a theory which describes the interaction of gravity with a Majorana
gravitino ${\Psi}_A$ and a three-index gauge field $A_{LMP}$.
The theory has several continuous symmetries: Local $N = 1$ SUSY,
$D = 11$ general covariance, Abelian gauge invariance for the three-form
$A_{LMP}$ and of course ${\mbox{SO}(10,1)}$ Lorentz invariance.  It also has
a discrete symmetry associated with the effect of spacetime reflections
on the gauge field.  This symmetry tells us [18] that the action 
and equations of motion in eleven dimensions are invariant under an
odd number of spatial (or temporal) reflections, together with the 
reversal of the sign of the gauge field:
\[
A_{LMP} ~{\longrightarrow}~ - A_{LMP}
\]
 
In fact, this discrete symmetry is {\it essential}
whenever we consider non-orientable spacetime manifolds in M-theory.  This is
because we typically think of the four-form $F_{LMNP}$ as being proportional
to some volume form, or anti-symmetric tensor ${\epsilon}_{LMNP}$.
It follows that on an non-orientable manifold, $F_{LMNP}$ will not
have a definite sign - the sign will change when we propagate around a 
non-orientable loop.  However, propagation around an orientation reversing
loop also reflects everything through an odd number of spacetime dimensions,
i.e., the equations of motion are still invariant even though the 
four-form is reduced to the status of a `pseudo-tensor'.  This means
that it still makes sense to talk about the eleven dimensional supergravity
equations of motion on non-orientable spacetimes.

Now, a key thing to notice is that it really is {\it not possible}
to consistently modify this structure in any way.  In particular, the 
Majorana condition for the gravitino is precisely what one needs in order
to match the number of bosonic and fermionic degrees of freedom.
One cannot just flippantly introduce other representations for the fermions.  

A pleasant feature of life in eleven dimensions is the fact that the real
Clifford algebra may be written as
\[
Cliff(10,1;{\Bbb R}) = {\Bbb R}(32)
\]

\noindent ${\Bbb R}(32)$ denotes the space of real 32$~{\times}~$32
matrices and $Cliff(10,1;{\Bbb R})$ denotes the set of objects ${\gamma}_{\nu}$
which satisfy the relation
\begin{equation}
{\gamma}_{\mu}{\gamma}_{\nu} + {\gamma}_{\nu}{\gamma}_{\mu} = +2g_{\mu\nu}
\end{equation}

\noindent where $g_{\mu\nu}$ is the metric on eleven dimensional Minkowski
space with the signature $(- + + + + + + + + + +)$.  In the usual way, these gamma
matrices act on a 32 dimensional space of Majorana spinors, 
which are real with respect to the relevant charge conjugation operator 
$C_{ij} = -C_{ji}$.  Explicitly, such a spinor is just a 32 component
column ${\psi}_k$, $k = 1, 2, 3, ... 32$.

It is {\it essential} for the contruction of eleven dimensional
supergravity that we are able to define, globally and consistently,
these Majorana fermions in any eleven dimensional spacetime we wish to consider.
Without such spinors, we can have no gravitino field with the right
number of degrees of freedom and similarly we cannot define generators
of supertranslations in superspace which will transform in the right way.

However, notice that these Majorana/SUSY conditions also select a {\it unique}
fermion type.  This is because, once we have made a choice for the
representations of $P$ and $T$, we are not allowed to introduce any
complex numbers (this would violate the Majorana condition) and we are
not allowed to do any parity doubling (then the fermions would have the
wrong number of degrees of freedom for SUSY).  But these are the only 
two mechanisms which we can use to generate other representations for $P$
and $T$!  In other words, there is always only one choice of $P$ and $T$
consistent with the Majorana/SUSY conditions in eleven dimensions.

It would therefore seem that the mathematical structure of M-theory
selects a unique pin bundle.  Four-dimensional multiplets,
the descendants of the unique eleven-dimensional structure, then inherit
this choice.  This elegant explanation of how nature 
may select a unique pin structure is
just another example of the power of M-theory

\vspace*{0.6cm}

{\noindent \bf Acknowledgements}\\

{The author would like to thank 
Prof. Gary Gibbons, Dr. Joshua Feinberg, and Prof. Robert Mann
for useful conversations.  This work was supported by NSF PHY94-07194
(at ITP, Santa Barbara) and by a Drapers Research Fellowship 
(at Pembroke College, University of
Cambridge). }\\
\vspace*{0.6cm}

{\noindent \bf References}\\

{\noindent [1] A.S. Wightman, {\it Superselection Rules; Old and New}, I1 Nuovo
Cimento, {\bf 110 B}, N. 5--6, Maggio-Giugno, (1995)}\\

{\noindent [2] R.F. Streater and A.S. Wightman, {\it PCT, Spin and Statistics, and
All That}, Addison-Wesley Publ. Co., (1964)}\\

{\noindent [3] Yakir Aharonov and Leonard Susskind, {\it Charge Superselection
Rule}, Phys. Rev. {\bf 155}, No. 5, (1967)}\\

{\noindent [4] E.P. Wigner, {\it Group Theoretical Concepts and Methods in
Elementary Particle Physics}, (Ed. F. Gursey), Gordon and Breach (1962)}\\

{\noindent [5] J. Giesen, {\it On a Non-CP-Violating Electric Dipole Moment of
Elementary Particles}, hep-th/9410236 }\\

{\noindent [6] L. D\c{a}browski, {\it Group Actions on Spinors}, Monographs and
Textbooks
in Physical Science, Bibliopolis, (1988)}\\

{\noindent [7] J.J. Sakurai, {\it Invariance Principles and Elementary
Particles},
Princeton Univ. Press, Princeton (1964)}\\

{\noindent [8] Cecile DeWitt-Morette, Shang-Jr Gwo, and Eric Kramer, {\it Spin
or Pin?}, Rev. Modern Phys. (in press) (1997)}\\

{\noindent [9] W. Heisenberg, Z. Phys. {\bf 77}, 1 (1932); for an English 
translation, see D.M.Brink, {\it Nuclear Forces}, Elmsford, NY: Pergamon (1965)}\\

{\noindent [10] Cecile DeWitt-Morette and B.S. DeWitt, {\it Pin Group in Physics},
Phys. Rev. D {\bf 41}, 1901-1907 (1990)}\\

{\noindent [11] Particle Data Group, {\it Review of Particle Properties}, Phys.
Rev. D {\bf 50}, 1173-1826 (1994)}\\

{\noindent [12] J.K. Baird, P.D. Miller, W.B. Dress and Norman F. Ramsey,
{\it Improved upper limit of the electric dipole moment of the neutron}, 
Phys. Rev. {\bf 179}, 1285-1291 (1969)}\\

{\noindent [13] C.G. Callan, R. Dashen and D. Gross, Phys. Lett. {\bf 63B},
334 (1976)}\\

{\noindent [14] R. Jackiw and C. Rebbi, Phys. Rev. Lett. {\bf 37}, 172 (1976)}\\

{\noindent [15] Donald H. Perkins, {\it Introduction to High Energy Physics},
Addison-Wesley Series in Advanced Physics (1972)}\\

{\noindent [16] V. Baluni, {\it CP-nonconserving effects in quantum 
chromodynamics}, Phys. Rev. D{\bf 19}, 2227-2230 (1979)}\\

{\noindent [17] E.P. Wigner, {\it Group Theory and its Application to the
Quantum Mechanics of Atomic Spectra}, Academic Press, New York (1959)}\\

{\noindent [18] M.J. Duff, B.E.W. Nilsson and C.N. Pope, {\it Kaluza-Klein
Supergravity}, Phys. Rep. Vol. 130, numbers 1 \& 2 (1986)}\\

{\noindent [19] J.P. Elliot and P.G. Dawber, {\it Symmetry in Physics},
Vol. 2, Oxford University Press, New York (1979)}\\

{\noindent [20] Y. Shirokov, Sov. Phys. JETP, vol. 11, page 101 (1960)}\\

\end{document}